\documentclass[12pt]{article}
\usepackage{times}
\usepackage{geometry}
\usepackage[utf8]{inputenc}
\usepackage{enumitem,amssymb}
\usepackage[numbers,sort]{natbib}
\usepackage{graphics,graphicx}
\usepackage{color}
\usepackage[table]{xcolor}
\usepackage{colortbl} 
\usepackage{tabularx}
\usepackage{caption}
\usepackage{paralist} 
\usepackage[colorlinks=true, citecolor=blue, linkcolor=black, urlcolor=black]{hyperref}
\usepackage{ragged2e}
\newlist{thematicfjhkjf}{itemize}{8}
\usepackage{pifont}
\usepackage{wrapfig}
\usepackage{graphics,graphicx}
\usepackage{color}
\usepackage{hyperref}
\usepackage{caption}

\def\apjl{\rm ApJL}
\def\apjs{\rm ApJS}

\def\aap{\rm A\&A}

\begin{document}
\raggedright
\huge
Astro2020 Science White Paper \linebreak

Neutrinos, Cosmic Rays and the MeV Band \linebreak
\normalsize

\noindent \textbf{Thematic Areas:} \hspace*{60pt} $\square$ Planetary Systems \hspace*{10pt} $\square$ Star and Planet Formation \hspace*{20pt}\linebreak
$\square$ Formation and Evolution of Compact Objects \hspace*{31pt} 
$\square$ Cosmology and Fundamental Physics \linebreak
  $\square$  Stars and Stellar Evolution \hspace*{1pt} $\square$ Resolved Stellar Populations and their Environments \hspace*{40pt} \linebreak
  $\square$    Galaxy Evolution   \hspace*{45pt}           \makebox[0pt][l]{$\square$}\raisebox{.0ex}{\hspace{-0.05em}$\checkmark$} Multi-Messenger Astronomy and Astrophysics \hspace*{65pt} \linebreak
  
\textbf{Principal Authors:}
Roopesh Ojha (UMBC/NASA GSFC, USA; Roopesh.Ojha@nasa.gov; Phone: +13012862972), Haocheng Zhang (Purdue University, USA; phitlip2007@gmail.com), Matthias Kadler (University of Würzburg, Germany; matthias.kadler@astro.uni-wuerzburg.de), Naoko K. Neilson (Drexel University, USA; nn344@drexel.edu), Michael Kreter (North-West University, South Africa; michael@kreter.org)

\textbf{Co-authors:} J.~McEnery (NASA GSFC, USA), S.~Buson (University of Würzburg, Germany and UMBC, USA), R.~Caputo (NASA GSFC, USA), P.~Coppi (Yale University, USA), F.~D'Ammando (INAF-IRA Bologna, Italy), A.~De Angelis (INFN/INAF Padua, Italy), K.~Fang (Stanford University), D.~Giannios (Purdue University, USA), S.~Guiriec (George Washington University, USA), F.~Guo (Los Alamos National Lab, USA), J.~Kopp (CERN and Johannes Gutenberg University Mainz), F.~Krauss (University of Amsterdam, The Netherlands), H.~Li (Los Alamos National Lab, USA), M.~Meyer (KIPAC/Stanford University, USA), A.~Moiseev (NASA GSFC, USA), M.~Petropoulou (Princeton University, USA), C.~Prescod-Weinstein (University of New Hampshire, USA), B.~Rani (NASA GSFC, USA), C.~Shrader (NASA GSFC/Catholic University of America, USA), T.~Venters (NASA GSFC, USA), Z.~Wadiasingh (NASA GSFC, USA). 
 \linebreak

\textbf{Abstract:} The possible association of the blazar TXS 0506+056 with a high-energy neutrino detected by IceCube holds the tantalizing potential to answer three astrophysical questions:\\
1. Where do high-energy neutrinos originate? \\
2. Where are cosmic rays produced and accelerated? \\
3. What radiation mechanisms produce the high-energy $\gamma$-rays in blazars? \\
The MeV $\gamma$-ray band holds the key to these questions, because it is an excellent proxy for photo-hadronic processes in blazar jets, which also produce neutrino counterparts. Variability in MeV $\gamma$-rays sheds light on the physical conditions and mechanisms that take place in the particle acceleration sites in blazar jets. In addition, hadronic blazar models also predict a high level of polarization fraction in the MeV band, which can unambiguously distinguish the radiation mechanism. Future MeV missions with a large field of view, high sensitivity, and polarization capabilities will play a central role in multi-messenger astronomy, since pointed, high-resolution telescopes will follow neutrino alerts only when triggered by an all-sky instrument.

\pagebreak

\justifying

\section{Introduction}

Neutrinos are uniquely efficacious probes because their low cross section and neutrality allow them to travel virtually unhindered through the Cosmos. They are unlikely to be attenuated in contrast to gamma-ray photons, which are subject to $\gamma\gamma$ interactions. Thus they are able to carry information about regions from which photons cannot escape, including some of the most compact as well as the most distant objects and environments in the Universe. In contrast to cosmic rays - which are nuclei mostly of hydrogen - neutrino propagation is not altered by magnetic fields and thus their origin can potentially be determined.  

Cosmic rays can reach energies up to $10^8$~TeV ($\gtrsim 1~\rm{EeV}$ is often called ultra-high-energy cosmic rays, UHECRs), orders of magnitude higher than what can be achieved with any particle accelerator on Earth. For over a century, the sources and underlying processes that accelerate cosmic rays have remained a major mystery. High-energy neutrinos are expected if cosmic rays entrained in (for example) a blazar jet interact with particles and radiation. In this sense, neutrinos are not just a messenger but perhaps the message itself. 

The detection of astrophysical high-energy neutrinos by the IceCube detector located at the South Pole has opened up an era of multi-messenger astrophysics and a number of candidate counterparts have been proposed \citep{Ahlers2015}. Blazars, a subclass of AGN whose jets are directed very close to our line of sight with violent $\gamma$-ray variability, have long been of interest as possible sources of UHECRs and high-energy neutrinos. The recent possible ($\sim3\sigma$) association of the blazar TXS\,0506$+$056 \citep{ice18a} with a high-energy neutrino lends support to this possibility that relativistic blazar jets may be the source of gamma rays, neutrinos, and cosmic rays.

Observations across the electromagnetic spectrum are crucial to identify and characterize the sources of neutrinos and cosmic rays. The MeV band is particularly salient because the MeV flux is the best proxy for the neutrino flux, as GeV-TeV $\gamma$-rays may be opaque within the source and at large distances due to attenuation. The brightest and most powerful blazars (such as flat-spectrum radio quasars, FSRQs) tend to have their peak emission in the MeV band, making it the most efficient band to probe radiation and particle acceleration in blazars. Polarization signals in the MeV band can distinguish between blazar emission models. Furthermore, current theories suggest that the MeV band is the most important band to constrain the neutrino production through the accompanying cascading pair synchrotron counterpart. Nonetheless, observationally MeV $\gamma$-rays are so far largely unexplored. To understand both electromagnetic and neutrino signatures from AGN jets, often referred to as the multi-messenger approach, \emph{support of observational and theoretical studies of $\gamma$-ray emission will be essential.}

\section{Centrality of the MeV band for neutrino--blazar studies}
Several attributes of the MeV band make it perfect to look for and characterize blazar counterparts to high-energy neutrinos. 
Triggered by the detection of individual high-energy IceCube neutrinos in coincidence with gamma-ray blazars in outburst \citep{Kadler2016,ice18a}, hadronic emission models have experienced a renaissance. These models involve interactions of high-energy protons in the jet with source-internal or external photons \citep{mannheim93,Reimer18,mastetal13,aharonian00}, and can explain the high-energy gamma-ray emission of blazars. 
There is general agreement that the broadband SED of the candidate neutrino blazars TXS\,0506+056 and PKS\,1424$-$418 can be explained with a leptohadronic model in which the leptonic component dominates the GeV band and the hadronic component leads to cascading emission that gets eventually released in the MeV and hard X-ray regime \citep{Gao17,Keivani18,Murase18,Gao19,Reimer18}. The down-cascading of gamma-ray photons below the GeV range may explain the mysterious neutrino excess (designated the `neutrino flare')  of $\sim 13$ neutrino events detected by IceCube
between September 2014 and March 2015, which has been independently
\begin{wrapfigure}{r}{0.65\textwidth}
\centering
\vspace{-0.2cm}
\includegraphics[width=0.65\textwidth]{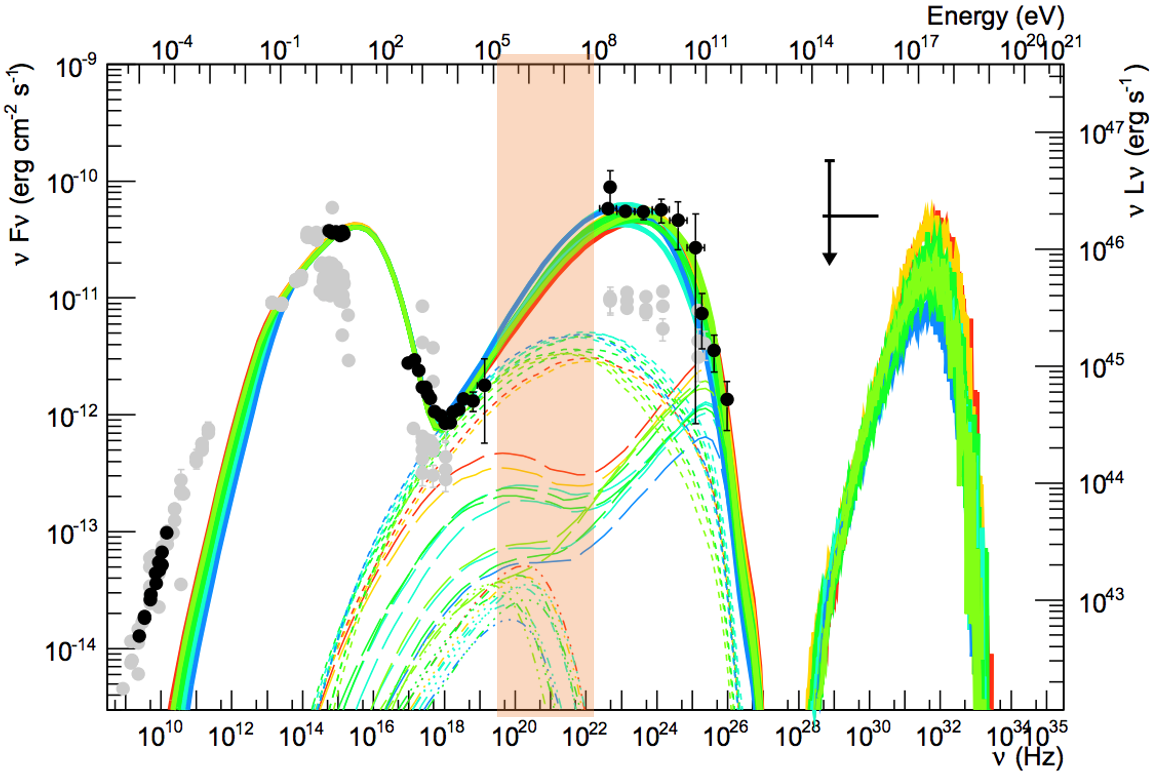}
\includegraphics[width=0.65\textwidth]{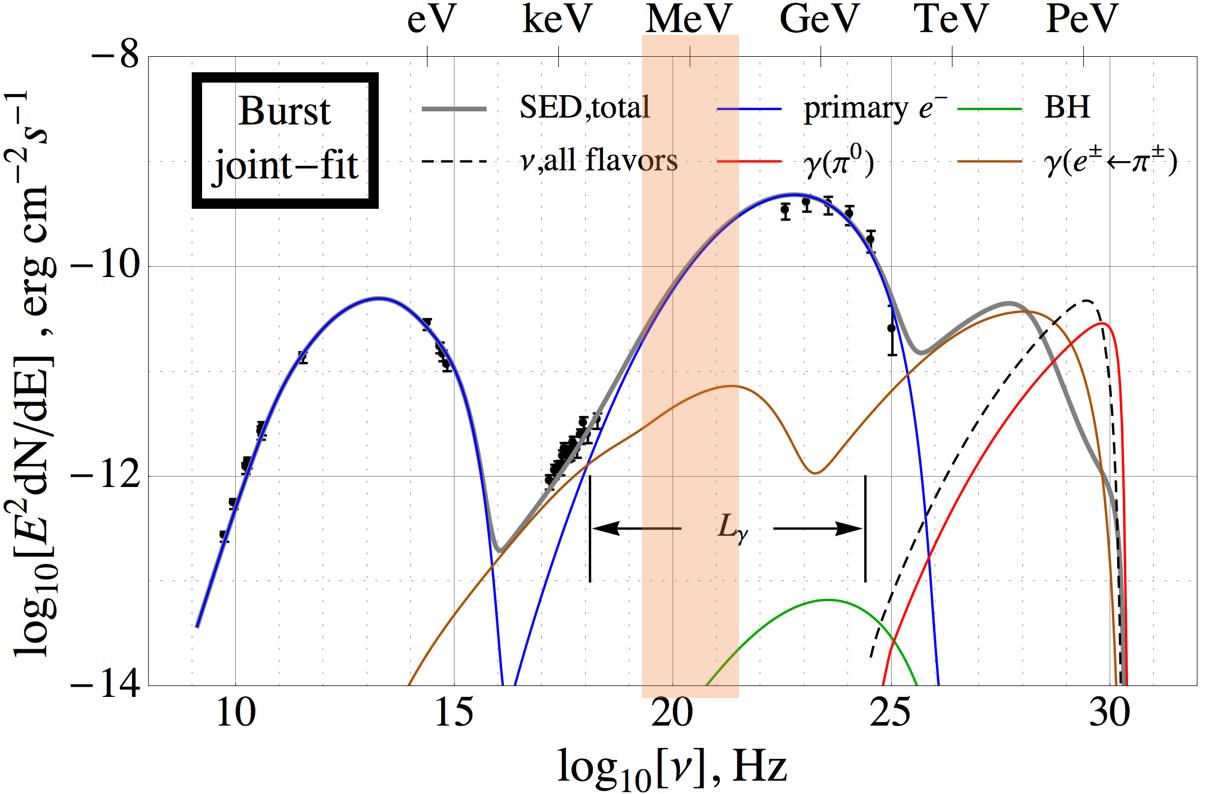}
\vspace{-0.2cm}
\captionsetup{font=footnotesize}
\caption{Top: Spectral Energy Distribution of TXS\,0506+056 and leptohadronic model \cite{Cerruti2019}. Bottom: Leptohadronic model of PKS~B1424$-$418 \cite{Gao17}. The highlighted MeV bands are essential to determine the cascading pair synchrotron contribution from hadronic interactions.\label{fig:SED}}
\vspace{-0.6 cm}
\end{wrapfigure}
associated with $3 \sigma$ significance with TXS\,0506+056 during an extended GeV-faint state \citep{Icecube18b}.
In these models, the MeV band is the key to finding the electromagnetic counterparts of the neutrino-brightest blazars in the sky, as it marks the transition from pair synchrotron contribution to inverse Compton (Fig. \ref{fig:SED}). 
It is even possible that a population of relatively faint GeV objects
could contribute significantly to the diffuse extragalactic neutrino flux or could even dominate it without being recognized in present-day blazar-neutrino correlation searches, which are largely focusing on \textit{Fermi}LAT-detected sources. In fact, some of the radio-brightest blazars are indeed GeV-faint and cannot be associated with $\gamma$-ray counterparts, even after 9 years of \textit{Fermi}LAT integration \citep{Lister15}.
In addition to closing the most critical gap in the current blazar SED coverage, the large field of view of planned instruments, such as AMEGO, will allow localization of neutrino candidate-associations for a large fraction of all high-energy neutrino events of interest. This capability is  essential to direct the observations of telescopes in other bands which typically have a very small field of view.

The maximum possible neutrino production rate of an individual blazar can be calculated based on calorimetric arguments from its spectral energy distribution and the long-term fluence in the keV to GeV range \cite{Krauss2014}, \cite{Kadler2016}. 
The GeV gamma-ray flux as measured by LAT is only of limited predictive value for the expected neutrino rate, because of the different
 possible shapes of blazar SEDs ranging from rather broad SED shapes and flat GeV spectra in the case of high-peaked BL\,Lac objects to MeV-peaked SEDs and steep GeV spectra for low-peaked FSRQs. 
 
  \begin{wrapfigure}{r}{0.65\textwidth}
\centering
\vspace{-0.4 cm}
\includegraphics[width=0.65\textwidth]{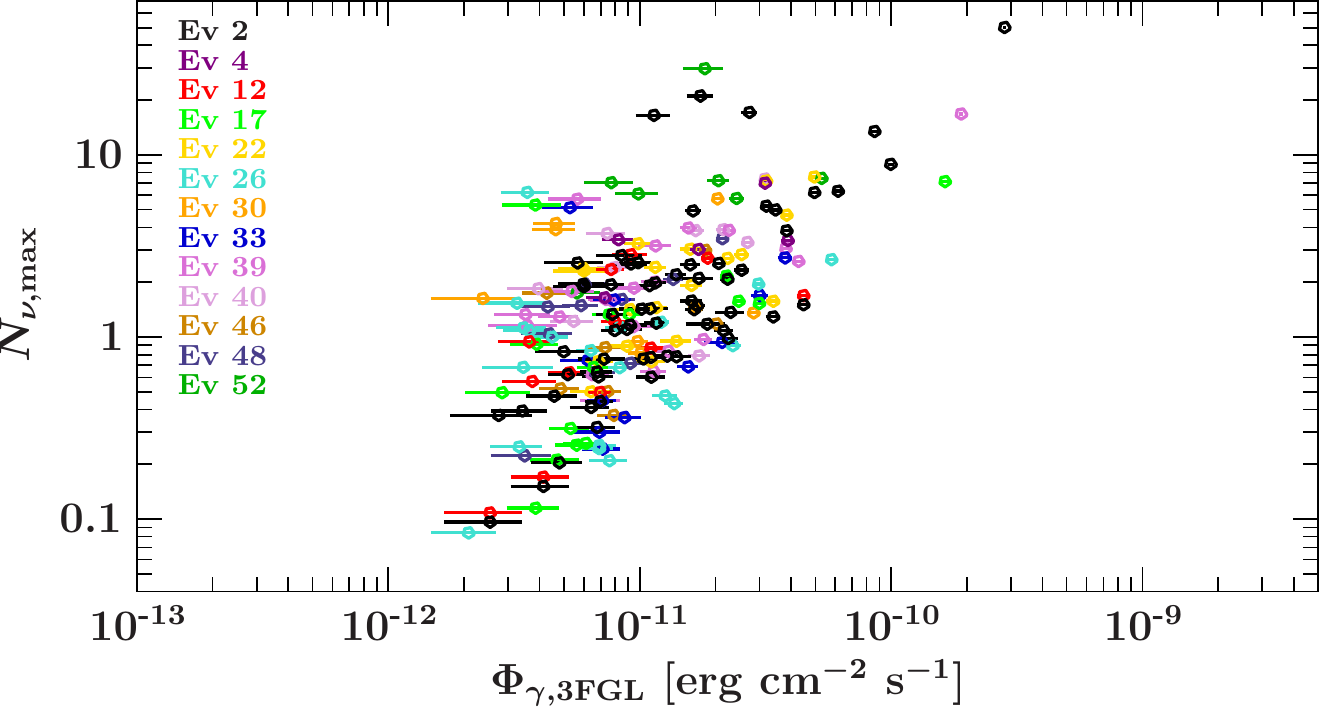}
\vspace{-0.8 cm}
\captionsetup{font=footnotesize}
\caption{Calculated maximum number of neutrinos vs. the given 3FGL flux for possible gamma-ray blazar counterparts to the
four-year IceCube HESE neutrinos above 100\,TeV and below 1\,PeV from \cite{Krauss2018}. The large scatter over more than one order of magnitude illustrates that the \textit{Fermi}LAT flux alone is not a good predictor
  of expected neutrino numbers. }
  \vspace{-0.4 cm}
\end{wrapfigure}

Neutrino estimates
using only the GeV fluxes can differ up to an order of magnitude from the real SED-based estimate \cite{Krauss2018}. The full high-energy calorimetric output can be determined much more accurately with measurements in the MeV band. This is particularly true for the most luminous flat-spectrum radio quasars, which tend to peak at MeV energies. Studies so far have found that the MeV part of the spectrum is the most important to characterize the most promising neutrino candidate blazars and to constrain the maximum possible neutrino production rate of the overall population of such sources \citep{Krauss2014,Krauss2018}. Poisson  probabilities  for  the  detection  of  single  or  multiple  high-energy neutrinos from individual high-fluence blazars are low, of the order of a few percent in the best cases \citep{Kadler2016}. Consistent with this, associations of neutrinos to individual blazars and results from stacked blazar samples of various characteristics are still very limited even after 10 years of IceCube observations \citep{Aartsen2017151}, \citep{Aartsen201745}. It is therefore mandatory to observe a large portion of the sky in the most crucial MeV energy band, in order to compile and characterize a statistical sample of reliable blazar-neutrino associations in the next decade.

\section{Centrality of MeV band in the search for sources of UHECRs}
The observed high polarization fraction (PF) in radio and optical bands has demonstrated that the blazar low-energy spectral hump is dominated by non-thermal electron synchrotron emission \citep{scarpa97}. The origin of the high-energy spectral hump from X-ray to $\gamma$-rays, however, has two competing scenarios. One scenario suggests that the same electrons that make the low-energy hump can inverse Compton (IC) scatter low-energy photons to X-rays and $\gamma$-rays. The low-energy photons may come from the low-energy synchrotron itself (synchrotron-self Compton, SSC), or external thermal photons (external Compton, EC) from accretion disk, broad line region, and dusty torus \citep[e.g.][]{maraschietal92, dermeretal92, sikoraetal94,bloommarscher96,ghisellinimadau96,boettcherdermer98}. As the two spectral humps are generally comparable, it requires that the seed photon energy density for the IC should also be similar to the magnetic field energy density that makes the low-energy synchrotron emission. Thus the IC scenario often implies a magnetic field strength of $\lesssim 1~\rm{G}$ \citep{boettcher13,cerruti15}. The alternative is the proton synchrotron (PS) scenario. In a high magnetic field ($\gtrsim 10~\rm{G}$), non-thermal protons can efficiently radiate through PS to make X-rays and $\gamma$-rays \citep[e.g.,][]{mannheim93,muecke03, mastetal13,aharonian00,mueckeprotheroe01}. To match the observed SED, the PS scenario generally requires the acceleration of UHECRs \citep{boettcher13,petrodimi15}. In addition to the two scenarios, there may exist a significant secondary synchrotron contribution in the X-ray to MeV $\gamma$-ray bands, originating from the hadronic interactions between non-thermal protons and the low-energy photon field (both low-energy synchrotron and external thermal emission). Hadronic interactions also produce high-energy neutrinos \citep{petromast15,Keivani18,Reimer18}. So far the two scenarios make similar SED fittings \citep{boettcher13}, thus \emph{we need additional constraints to pinpoint the radiation mechanisms in the high-energy spectral hump, and diagnose the radiating particles and magnetic field strength.} Furthermore, theoretical and numerical simulations have shown that both shock and magnetic reconnection can give rise to the non-thermal electrons and protons that make the radiation \citep{marscher85,spada01,larionov13,zhang16a,boettcher19,romanova92, giannios09, sironietal15, petroetal16, zhang18}. Which mechanism dominates the particle acceleration in AGN jets largely relies on the physical conditions of the acceleration sites, in particular the magnetic field strength and morphology. As the $\gamma$-rays often show fast variability that implies the most fierce particle acceleration, \emph{temporal $\gamma$-ray signatures are vital to probe the extreme particle acceleration in AGN jets.}

\begin{wrapfigure}{r}{0.6\textwidth}
\centering
\vspace{-0.5 cm}
\includegraphics[width=0.6\textwidth]{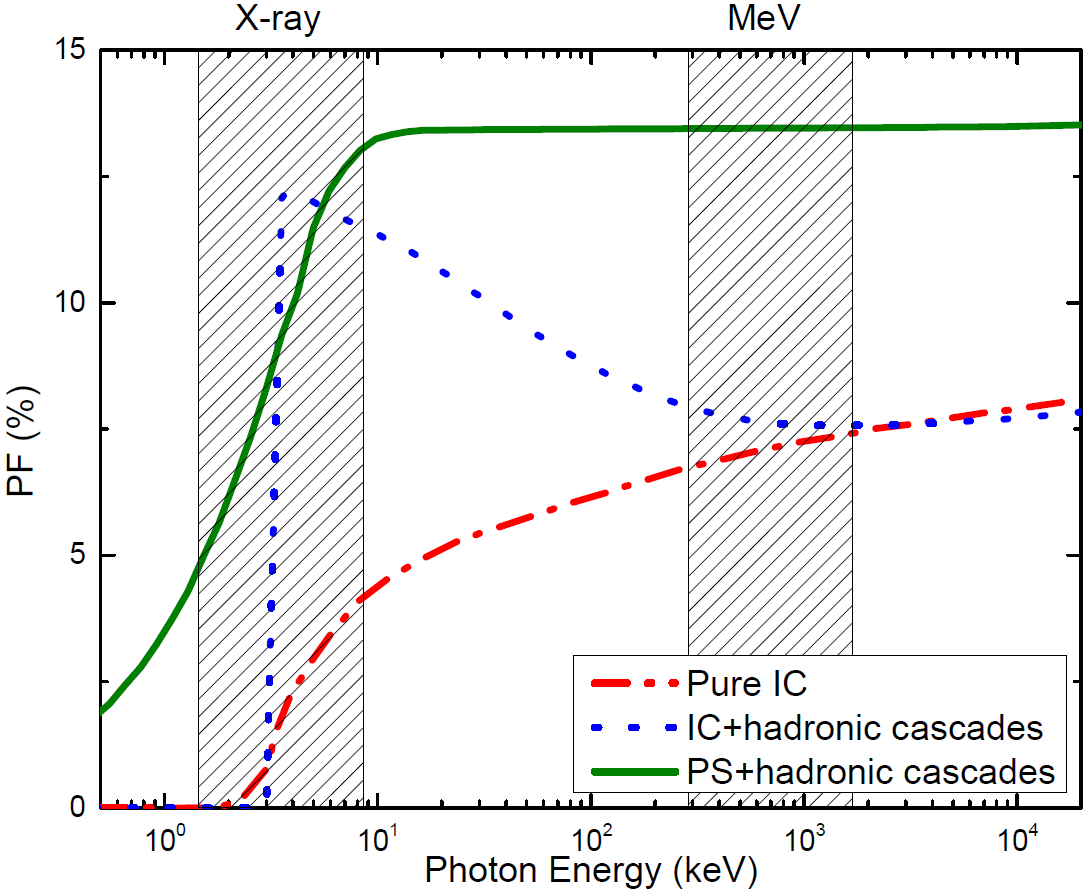}
\captionsetup{font=footnotesize}
\vspace{-0.8 cm}
\caption{MeV $\gamma$-ray polarization fraction as a unique diagnostic tool of inverse Compton and proton synchrotron scenarios of high-energy blazar emission. The PF calculation is based on detailed TXS~0506+056 spectral modeling \cite{Zhang19}. \label{fig:specpol}}
\vspace{-0.4 cm}
\end{wrapfigure}

The largely unexplored MeV $\gamma$-ray band is crucial to advance our knowledge in AGN jet physics. In an IC scenario, MeV bands often mark the spectral transition from SSC to EC, and if there is a secondary synchrotron component from hadronic interactions, it should be significant from X-ray to MeV bands as well \citep{boettcher13,cerruti15}. In a PS model, MeV bands can probe the contribution by the secondary synchrotron component, shedding light on the neutrino production. Therefore, both spectral shape and variability in MeV bands can put unprecedented constraints on the AGN jet radiation processes. A very interesting and novel opportunity is MeV polarimetry. Theoretical studies suggest that the IC and PS mechanisms result in very different PF \citep{zhang13,zhang16,paliya18}. Based on detailed spectral fitting of the recent TXS~0506+056 event \citep{Zhang19}, Fig. \ref{fig:specpol} demonstrates that \emph{MeV polarimetry can unambiguously distinguish the IC and PS scenarios by the PF}. With the aid of X-ray polarimetry and neutrino detection, we can diagnose the contributions of various radiation mechanisms in the high-energy spectral hump. In particular, if the PS scenario makes the high-energy hump, temporal MeV polarization signatures can probe the magnetic field strength and morphology evolution in the particle acceleration sites, providing novel constraints on the generation of UHECRs. \emph{Next-generation MeV (Compton and pair) instruments with great spectral and temporal sensitivity as well as polarimetry capability will be the best to advance our knowledge of AGN jet physics.}

The fast evolving physical conditions in the blazar emission region and the complicated hadronic processes therein, including the radiative transport, feedback on the non-thermal particles, and neutrino production, prevent the use of simple steady-state analytical models. In the recent decade, first-principle numerical simulations, including magnetohydrodynamic (MHD) and particle-in-cell (PIC) methods, have successfully revealed the time-dependent evolution of fluids and non-thermal particles in the blazar emission region \citep{Mizuno09,Spitkovsky08,Sironi09,Sironi14,Guo14}. Advanced Monte-Carlo and ray-tracing methods have been fruitful in reproducing multi-wavelength SEDs, light curves, and optical polarization signatures \citep{Chen14,Zhang15,marscher14}. These new developments have sparked the concept of multi-physics simulations, which aim to self-consistently connect fluid dynamics, particle acceleration, and radiative transfer (preliminary works include, \citep{tavecchio18,zhang18,christie19}). \emph{Multi-physics numerical simulations will be a top priority in the next decade to leverage the multi-messenger studies of AGN jets.}

\begin{wrapfigure}{r}{0.5\textwidth}
\centering
\includegraphics[width=8 cm]{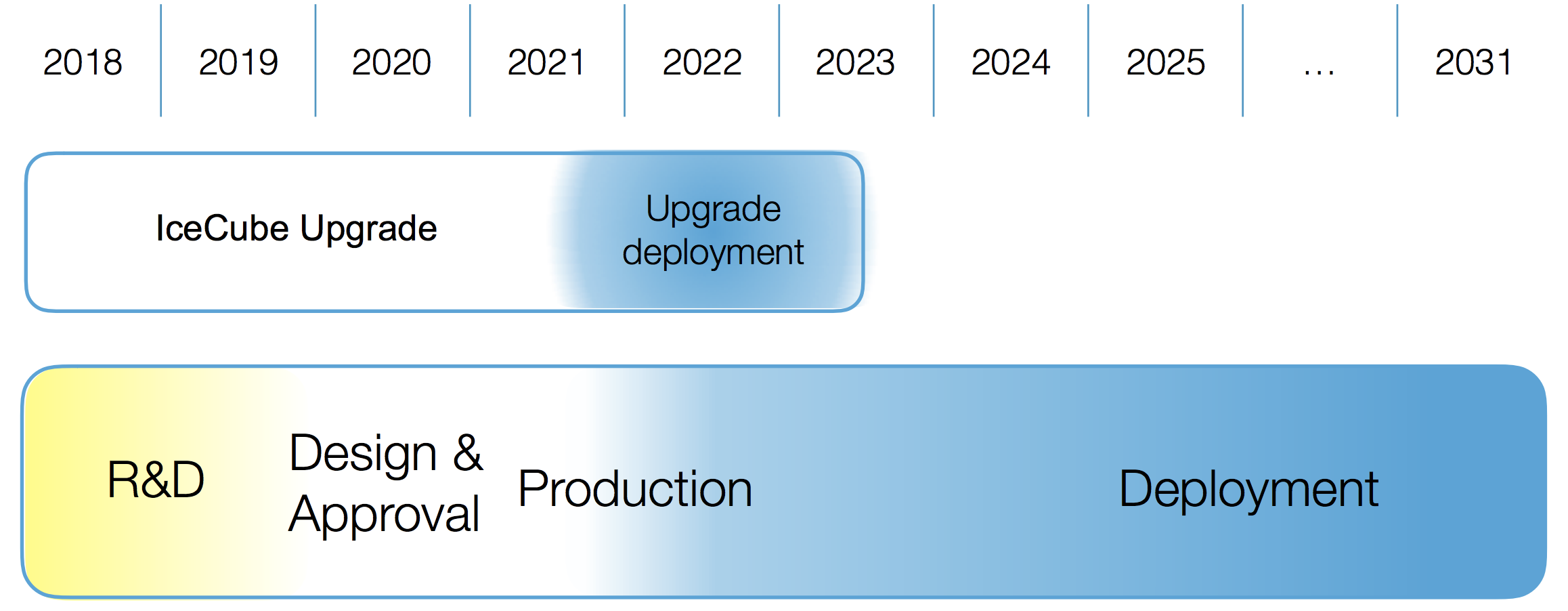}
\captionsetup{font=footnotesize}
\caption{IceCube-Gen2 Timeline}
\end{wrapfigure}

\section{MeV and Neutrino Capabilities Required in the Next Decade}
\subsection{MeV Detection} To achieve the goals discussed above, we need an MeV instrument with high sensitivity, good energy resolution, polarization capability, and a large field of view. For example, AMEGO will have a FOV $>$2.5 sr, covering the entire sky every 3 hours providing an all-sky instrument to match IceCube and other neutrino experiments. A joint detection gives much greater scientific returns with good localization and quick repointing making better background suppression and thus increasing sensitivity. Further, `pointed' telescopes at other bands are likely to only follow neutrino alerts if triggered by large FOV instruments in the MeV band. 

\subsection{IceCube and Neutrino detection capabilities}
The IceCube Collaboration has been approved to deploy seven additional strings of photon sensors in the deep, clear Antarctic ice at the bottom center of the existing detector by 2023, forming the IceCube Upgrade. Newly developed photon sensors and new calibration devices will allow IceCube to better model the optical properties of the ice, reducing systematic uncertainties and enhancing IceCube's already strong contribution to multi-messenger astrophysics via improved reconstruction of the direction of high-energy cascade events for point source searches and enhanced identification of PeV-scale tau neutrinos. This project also provides transitions for construction and detector developments to a proposed IceCube-Gen2 Facility, aimed to be deployed around 2030. Gen2 is a unique, multi-component cosmic neutrino observatory that expands IceCube's existing energy reach by several orders of magnitude in both directions. An order of magnitude more astrophysical neutrinos are expected to be detected by the expanded in-ice portion of the detector with surface detector components that veto atmospheric background particles efficiently.

In addition to IceCube, several other neutrino detectors, the ARCA instrument on KM3Net, Baikal-GVD, ARA, and GRAND, are expected to be operational in the next decade. Located on different hemispheres and with complementary capabilities, much superior to the already highly successful past decade, the next decade will see a true all-sky coverage of neutrino telescopes. For cosmic rays, AugerPrime and TAx4 plan to operate in the next decade. These will enhance the number of astrophysical sources of interest to be followed up at $\gamma$-rays, and create a better picture of neutrinos, cosmic rays, and the MeV band.

\pagebreak

\bibliography{NeutrinoCR.bib}

\end{document}